\documentclass[12pt]{article}

\input epsf
\def\t{\widetilde}
\def\beq{\begin{eqnarray}}
\def\eeq{\end{eqnarray}}

\def\mpl{M_{\rm Pl}}

\def\lsim{\mathrel{\rlap{\lower3pt\hbox{\hskip0pt$\sim$}}
     \raise1pt\hbox{$<$}}}         
\def\gsim{\mathrel{\rlap{\lower4pt\hbox{\hskip1pt$\sim$}}
     \raise1pt\hbox{$>$}}}         

\begin{document}


\begin{flushright}
{ NYU-TH-05/08/15}
\end{flushright}
\vskip 0.9cm

\centerline{\Large \bf 
Short Distance Non-perturbative Effects of }
\vspace{0.5cm}
\centerline{\Large \bf Large Distance Modified Gravity  }

\vskip 0.7cm
\centerline{\large Gregory Gabadadze and Alberto Iglesias}
\vskip 0.3cm
\centerline{\em Center for Cosmology and Particle Physics}
\centerline{\em Department of Physics, New York University, New York, 
NY, 10003, USA}

\vskip 1.9cm 

\begin{abstract}

In a model of large distance modified gravity 
we compare the non\-perturba\-tive Schwarzschild solution of 
hep-th/0407049 to approximate solutions obtained previously. 
In the regions where there is a good qualitative 
agreement between the two, the nonperturbative  
solution yields effects that could have 
observational significance. These effects reduce, by a factor of 
a few, the predictions for the additional precession of the 
orbits in the Solar system, still rendering them in an 
observationally interesting range. The very same effects  
lead to a mild anomalous scaling of the additional scale-invariant 
precession rate found by Lue and Starkman.

\end{abstract}

\newpage

\section{Introduction}

The DGP model of large distance modified gravity \cite {DGP}
has one adjustable parameter -- the distance scale $r_c$. 
Distributions of matter and  radiation which are homogeneous and 
isotropic at scales $ \gsim r_c$ exhibit in this model the 
following properties: for distance/time 
scales $\ll r_c$ the solutions approximate General Relativity (GR)
to a  high accuracy, while for scales $\gsim r_c$
they dramatically differ \cite {DGP,Deffayet,DDG,DDGV}. 
Postulating that $r^{-1}_c\sim H_0\sim 10^{-42} {\rm GeV}$
the deviations from GR could lead to interesting observational  
consequences in late-time cosmology, see, e.g., 
\cite {DDG,Landau}, \cite{Lue2}--\cite{Spergel}.

On the other hand, sources of matter and radiation 
with typical inhomogeneity scale  less than $r_c$
have somewhat different properties. 
These are easier to discuss  for  a  Schwarzschild source --
a spherically-symmetric distribution of matter of the mass $M$ 
and radius $r_0$, such that $r_M < r_0  \ll r_c$ 
($r_M\equiv 2G_NM$  is the  Schwarzschild  radius 
and $G_N$ the Newton constant).  For such a source a 
new scale, that is a combinations of $r_c$ and $r_M$, emerges 
(the so-called  Vainshtein scale\footnote{A similar, but not 
exactly the same scale was discovered by Vainshtein in massive 
gravity \cite {Arkady}, hence the name.}) \cite {DDGV}: 
\beq
r_*\equiv (r_M r_c^2)^{1/3}\,.
\label{r*}
\eeq
Above this scale gravity of a compact object deviates 
substantially from the GR result. Note that $r_*$ is huge for 
typical astrophysical objects. An isolated star of 
a solar mass would have $r_* \sim 100~ pc$. However,  
if we draw a sphere of a $100~pc$ radius   
with the Sun  in its center  there will be many other starts
enclosed by that sphere.  The matter enclosed
by this sphere would have even larger  $r_*$. 
We could  draw a bigger sphere, but it will enclose 
more matter which would  yield yet larger  $r_*$ and so on.  
An isolated object which  could  be separated from  
a neighboring  one  by a distance larger than its own $r_*$
is a cluster of galaxies. For typical clusters,  $r_*\sim (few~Mpc)$
is just somewhat larger than their size and is smaller than 
their average separation. The above arguments suggest that interactions of 
isolated clusters will be different in the DGP model.  On the other hand,  
at scales beneath a few Mpc or so, there will be agreement with the GR 
results with potentially interesting small deviations.  
Below we discuss  these issues in detail on  an example 
of a single isolated  Schwarzschild source. There exist 
in the literature two different solutions for the Schwarzschild 
problem in the DGP model.  The first one is based on approximate 
expansions in the $r \ll r_*$ and $r \gg r_*$ regions \cite 
{DGP,DDGV,Gruzinov} (see also \cite {Lue1,Porrati}). 
We call this set of results the perturbative  Schwarzschild (PS) solution. 
The second one \cite {GI} is a solution that interpolates smoothly 
from $r\ll r_*$  to $r\gg r_c\gg r_*$,  and is non-analytic in the 
either parameters  used to obtain the PS solution.  
We call this the non-perturbative 
Schwarzschild (NPS) solution. It is important to understand which of 
these two solutions, if any, is physically viable. Since neither of the two 
can be solved completely without numerical simulations, 
a first step to discriminate between them
would be to look closely at the theoretical 
differences, as well as predictions that could by tested 
observationally. This is the goal of the present note.

\section{Qualitative discussions} 

We will study separately  two regimes,  $r \ll r_*$ and $r \gg r_*$.

{\bf \underline {(I) $r \ll r_*$}}. In this regime 
the standard  $G_N$ expansion breaks down \cite {DDGV}. 
How could  one proceed? One way is to perform an expansion 
in powers of  $m_c=r_c^{-1}$ \cite {DDGV}. This expansion 
breaks down above $r \sim r_*$  but is well suited
for the $r\ll r_*$ domain (Kaloper \cite {Kaloper}
recently used a different expansion. His proposal 
could prove  to be useful for a broad class 
of problems). A Schwarzschild metric in  the small $m_c$  
expansion  was calculated  by Gruzinov \cite {Gruzinov} 
(see also \cite{Lue1}). It is instructive to compare the 
result of \cite {Gruzinov} with the NPS solution of \cite {GI}. 

Let us  start with  the  Newton  
potential $\phi(r)$. The expansion of the exact result 
of \cite {GI} for $r\ll r_*$ leads:
\beq
-2\phi = {r_M\over r}-\alpha m_c^2r^2 
\left({r_*\over r}\right)^{{3\over 2} - \beta }+\dots~,
\label{potential}
\eeq
where $\beta = 3/2 - 2(\sqrt{3}-1)\simeq 0.04$, and      
$\alpha$ is a number to be discussed in detail below. 
The above result, but with $\beta =0$, is what was 
first obtained in a  small $m_c$ expansion 
\cite {Gruzinov}. The NPS solution of \cite {GI} gives 
$\beta \simeq 0.04$, it depends on irrational powers of 
$m_c$ \cite {GI}, and it differs by that from the small $m_c$ 
expansion results.

Is the above difference important?
As was demonstrated in Refs. \cite{Lue1} and \cite {DGZ},
the modification of the Newton potential in  (\ref {potential}),
although tiny, could lead to a measurable  precession 
of orbits in the solar system (see, Refs. \cite {Iorio1} for further 
studies). The above works used the potentials 
obtained in the small $m_c$ expansion, e.g., used 
(\ref {potential}) with $\beta=0$.  Although 
$\beta$ is tiny,  the ratio $(r_*/r)$ is typically huge in the cases of 
interest,  therefore, taking into account the effects of a nonzero $\beta$ 
could lead to  appreciable differences in  the predictions of 
the PS and NPS solutions. We will study this issue in the next 
section.

Consequences of the modified potential  (\ref {potential}) 
could  be understood as well in terms of invariant curvatures. 
The Schwarzschild solution in GR has 
zero scalar curvature. In contrast with this, 
the solution (\ref {potential}) generates a nonzero Ricci scalar 
that extends to $r\sim r_*$ in the NPS solution 
(see, \cite {GI} and discussions below).
This can be seen by looking at the trace equation in the DGP model:
\beq
R-3m_c K = T \,,
\label{curvature}
\eeq
where $R$ is the 4D Ricci scalar, $K$ is a trace of an extrinsic curvature 
and  $T$ is a trace of the stress-tensor times $8\pi G_N$ (for the 
ADM formalism in the DGP model see, e.g., \cite {Dick,CedricMourad}). 
This has to be compared with the trace equation in
GR: $R=T$. The second term on the LHS of (\ref {curvature})
is not zero outside the source and, therefore, gives rise to 
nonzero $R$. This curvature, although tiny,  extends  to enormous 
scales of  the order of $r\sim r_*$ \cite {GI}. The sign of the 
curvature depends on a choice of the boundary conditions in the bulk, 
since the latter determines the sign of $K$. There are two choices for 
this. The so-called conventional branch corresponds to  
a negative (AdS like)  curvature produced by the Schwarzschild  source,
while  the selfaccelerated branch \cite {Deffayet} 
corresponds to a  positive (dS like) $R$.  This is 
reflected in the sign of the coefficient $\alpha$ in (\ref {potential})
which takes a positive value on the conventional branch 
and becomes negative  on the selfaccelerated branch: $\alpha \simeq 
\pm 0.84$. Therefore, there is 
an additional tiny attraction toward the source on the conventional 
branch and a repulsion of the same magnitude on the 
selfaccelerated branch. This change of sign was first 
found  by  Lue and Starkman  \cite {Lue1} in the context of the 
PS solution.

{\bf \underline {(II) $r\gg r_*$}}.  In this regime the 
small $m_c$ expansion breaks down. However, the conventional 
$G_N$ expansion can be readily used  \cite {DGP,DDGV}. 
The results are \cite{DGP}: 

(A) For $r\gg r_*$
DGP gravity is a tensor-scalar theory, where the extra 
scalar couples to matter with the gravitational strength:
the vDVZ phenomenon \cite {Veltman,Zakharov}.  

(B) The Newton potential scales as $1/r$ for $r_* \ll r\ll r_c$
which smoothly transitions into the $1/r^2$ potential at
$r\gg r_c$. 

These properties of the PS solution 
were reconfirmed in detailed studies of  Refs. 
\cite {Gruzinov,Lue1,Porrati,Siopsis,Tanaka}.
Could the PS  solution interpolate from 
$r \ll r_*$ to $r \gg r_c$? The above question is  
related to the  following one:  what is a gravitational 
mass that is felt by an object  separated from the source  
at a distance $r\gg r_*$? The PS solution implies that this is just 
the bare mass $M$ of the original source.
On the other hand, one may expect that 
the curvature created by the source in the domain
$r\ll r_*$ would also contribute to this effective mass
(the ADM mass) \cite {GI}. If so, unless there is 
a hidden nontrivial cancellation, a  putative observer
at $r\gg r_*$  would  measure  an effective mass 
different from $M$. The above property is captured by the  
NPS solution of  Ref. \cite  {GI}. It has the following features: 

(A$^\prime$) For $r\gg r_*$ it is a solution of a tensor-scalar
gravity (as in (A) above); 

(B$^\prime$) The Newton potential scales as $1/r^2$
for $r \gg r_*$ (different from (B)).
 
An attractive feature of the  NPS solution is that it 
smoothly interpolates from  $r\ll r_*$ to $r>>r_c$.  
However, a somewhat unusual fact is that it does not recover the 
results of the $G_N$ expansion. This will be discussed  in  the 
reminder of this section (readers who are not interested in these 
somewhat technical issues could  directly go to the 
next section without loss of clarity).

Why is that, that the  NPS solution \cite {GI} 
does not agree with the results of the perturbative $G_N$
expansion, even in  the regime $r\gg r_*$, 
where the latter approximation  is internally 
self-consistent?  There could be a few different 
reasons for this. Formally, one is solving 
nonlinear partial differential equations  and these can have 
different solutions even with the same boundary conditions.
In our two cases, however, the boundary conditions are 
somewhat different:  the PS solution is supposed
to describe  the same mass $M$ at short and large distances,
while the NPS solution  matches $M$ at the short scales
but asymptotes to a screened mass at the large scales
\footnote{The boundary conditions at the brane 
are also different, see a footnote on page 6.}. Then either the PS and NPS 
solutions belong to different sectors and are both stable, 
or at least one of them should be unstable. 
In the former case, one should distinguish between them observationally, 
while in the latter case a relevant point  would be that 
the ADM mass of the NPS  solution is smaller \cite {GI}.
In a very qualitative way, this can be understood as follows.
A deviation from the conventional metric at  $r\ll r_*$ 
sclaes as $m_c \sqrt{r_M r}$ (we ignore small $\beta$ here.)
This can give rise to a scaling of the scalar curvature 
$m_c\sqrt{r_M}r^{-3/2}$. The curvature extends roughly to 
$r \sim r_*$, and the integrated curvature scales as 
$m_c \sqrt{r_M} r_*^{3/2}\sim r_M$. Then, the "effective mass"'
due to this curvature can be estimated as $r_M \mpl^2 \sim M$, 
which is of the order of the mass itself.
  
On the other hand, it may well be that there is a certain 
``discontinuity'' between  the linearized and full 
non-linear versions of the DGP model in 5D. This could 
result from  a different number of constraints  one has 
to satisfy depending on 
whether solutions are looked for in the linearized 
approximation or in the full non-linear 
theory.  For instance,  one of the bulk equations 
can be combined with the junction condition 
in 4D to yield:
\beq
3 m_c^2 R = { R^2 - 3 R^2_{\mu\nu} }\,.
\label{RR}
\eeq
On a flat background  both terms on the RHS of 
(\ref {RR}) contain  at least quadratic terms in the fields. 
Therefore, according to (\ref {RR}), $R$ has to be zero
in the linearized approximation.
The latter condition happens to be  a consequence
of the other linearized equations of the theory as well; 
therefore, (\ref {RR}) is  trivially satisfied as long as 
those other equations are fulfilled.
This changes at the nonlinear-level:
Eq. (\ref {RR}) becomes an additional constraint 
that one has to satisfy on top of the other equations.
Because of this: (i) The solutions of the linearized theory may not be 
supported by the nonlinear equations (the phenomenon known as 
"linearization instability" in gravity).  
(ii) New non-perturbative solutions that do not exist 
in the linearized theory may emerge. One way to decide on 
the point  (i),  is to study  solutions for other 
sources and see whether a similar phenomenon  takes place. 
The NPS solution of \cite {GI} is an 
explicit example of the point (ii).


\section{Explicit solution}

{}We consider the action of the DGP model \cite{DGP}: 
\begin{equation}
 S= M_*^3\int d^5x \sqrt{-g} R+M_P^2\int d^4x\sqrt{-\t g}\t R~.
 \label{action}
\end{equation}
 Here, the $(4+1)$ coordinates are $x^M=(x^\mu,\ y)$, $\mu=0,
\dots, 3$ and $g$ and $R$ are the determinant and curvature of 
the 5 dimensional metric $g_{MN}$,
while $\t g$ and $\t R$ are the determinant and curvatures of the 4 dimensional
metric $\t g_{\mu\nu}=g_{\mu\nu}(x^\mu,y=0)$. The Gibbons-Hawking 
\cite {GH} surface term that guaranties correct equations of motion is 
implied in the action (\ref {action}). $M_P$ denotes the 4D  Planck
mass and is fixed by the Newton constant. On the other hand, the 
scale $M_*$ is traded for the parameter $r_c \equiv  M^2_p/2 M^3_*$
discussed in the previous section.

{}The NPS solution studied in \cite{GI} 
is found by considering a static metric with spherical symmetry on the
brane and with ${\bf Z}_2$ symmetric line element:
\begin{equation}
ds^2=-{\rm e}^{-\lambda} dt^2+{\rm e}^\lambda dr^2+r^2d\Omega^2+\gamma\ dr
dy+{\rm e}^\sigma dy^2~,\label{ds}
\label{interval}
\end{equation}    
where $\lambda,\ \gamma,\ \sigma$ are functions of 
$r=\sqrt{x^\mu  g_{\mu \nu} x^\nu}$ and $y$. The ${\bf Z}_2$ symmetry 
across the brane implies that $\gamma$ is 
an odd function of $y$ while the rest are even. The brane is chosen 
to be straight in the above coordinate system\footnote{One could 
transform (\ref {interval}) to the coordinate system where the metric 
is diagonal $ds^2 =-A(r,z)dt^2 + B(r,z)d\rho ^2 + C(r,z) d\Omega^2 +dz^2$, and 
$A\neq B$. In this system our brane will be bent.}.

The exact solution for $y\to 0+$ is given implicitly as follows:
\beq
{\rm e}^{-\lambda} = 1-{P(r)\over r}~,
\eeq
where $P$ is obtained from
\beq
P(r)=-{3\over 2}m_c^2 \int {\rm d}r\ r^2 U(r)~, 
\label{P}
\eeq
in which $U$ can have two different behaviors corresponding to the solution 
of the following two equations (giving rise to a conventional  and 
selfaccelerated branch respectively): 
\begin{eqnarray}
(k_1 r)^8&=&-{(1+3U+f)\over U^2(3+3U+\sqrt{3}f)^{2\sqrt{3}}(-5-3U+f)}~,
\label{sol1}\\
(k_2 r)^8&=&-{(-5-3U+f)(-3-3U-\sqrt{3}f)^{2\sqrt{3}}\over (U+2)^2 
(1+3U+f)}~,\label{sol2}
\end{eqnarray}
where $f=\sqrt{1+6U+3U^2}$ and $k$ is an integration constant. 

{}Note that in this parametrization the gravitational potential $\phi$ in weak 
field approximation is easily obtained, namely 
\beq
\phi\equiv - {P(r)\over 2r}~. 
\eeq
The off-diagonal metric component, $\gamma$, is determined from 
\beq
{2r^2P_r\over P_{rr}}={\left(r^4\gamma {\rm e}^{-\lambda}\right)_r
\over \left(r\gamma {\rm e}^{-\lambda}\right)_r}~,
\eeq
and the $yy$ component from
\beq
{\rm e}^\sigma =m_c^2\left[{\left(r^4\gamma {\rm e}^{-\lambda}\right)_r\over
2r^2P_r}\right]^2+{\rm e}^{-\lambda}\gamma^2~.
\eeq
The profile $\lambda_y$ for $y\to 0+$ can be computed as well:
\beq
\lambda_y= {\rm e}^{-\lambda}\gamma_r~.
\eeq
The two integration constants, $k$ and the one produced in the integration 
(\ref{P}), are determined by imposing appropriate boundary conditions at the 
source (namely, $P(r\to 0+)\to r_M$) and at large distances,  
(namely, $\lambda\sim \t r_M^2/r^2$  in the conventional branch or 
$\lambda \sim m_c^2r^2+\t r_M^2/r^2$ in the selfaccelerated branch 
and no $1/r$ term). 


\subsection{Conventional branch}

{}The conventional branch is obtained from the solution of (\ref{sol1}).
As shown in \cite{GI} the boundary conditions ($P(0)=r_M$, $P(+\infty)=0$) 
determine the exact relation between $k_1$ and $r_*$, namely
\beq\label{k1}
2(r_*k)^3=c~,
\eeq
where $c$ is the following integral:
\beq\label{int}
c=\int_0^\infty
\left[-{(1+3U+f)\over U^2(3+3U+\sqrt{3}f)^{2\sqrt{3}}(-5-3U+f)}\right]^{3/8}
{\rm d}U\approx 0.43~.
\eeq
The solution has the following 
asymptotic behavior. At large distances, $r\gg r_*$ ($U\to 0^+$), we obtain
\beq
{P(r)\over r} ={\t r_{M_1}^2\over r^2}+\dots~,
\eeq
where,
\beq
\t r_{M_1}^2={3\sqrt{2}\over 4(3+\sqrt{3})^{\sqrt{3}}} {m_c^2\over k_1^4}
\approx 0.56\ r_M r_*~,
\eeq
while at short distances, $r\ll r_*$ ($U\to +\infty$), we get
\beq\label{short}
{P(r)\over r}={r_M\over r}-\alpha_1 m_c^2r^2 
\left({r_*\over r}\right)^{2(\sqrt{3}-1)}+\dots~,
\eeq
where
\beq
\alpha_1=6^{(\sqrt{3}-1)/2}{1+\sqrt{3}\over 4(3\sqrt{3}-1)}
\left(3+\sqrt{3}\over3-
\sqrt{3}\right)^{(\sqrt{3}-1)/4}(k_1r_*)^{2(1-\sqrt{3})}\approx 0.84~.
\eeq
As we see, a short distance observer at $r_M \ll r\ll r_*$ 
would measure the gravitational mass $M$ with a small corrections
to Newton's potential, while the large distance observer
at $r\gg r_*$  would measure an effective gravitational mass
$ \sim M(r_M/r_c)^{1/3}$ \cite {GI}. The latter 
includes the effects of the 4D curvature.


\subsection{Selfaccelerated branch}

{}The solution on the selfaccelerated branch is obtained from (\ref{sol2}).
The relation between $k_1$ and $r_*$ is obtained, as in the conventional case,
by imposing boundary conditions ($P(0)=r_M$, 
$P(r)-m_c^2r^3\to 0$ for large $r$). This gives
\beq
2(r_*k_2)^3&=&-\int_{-\infty}^{-2}
(U+2){{\rm d}\over {\rm d}U}
\left[-{(1+3U+f)\over U^2(3+3U+\sqrt{3}f)^{2\sqrt{3}}(-5-3U+f)}\right]^{3/8}
{\rm d}U\nonumber\\
&=&6^{3\sqrt{3}/4}c\approx 4.41~.\label{k2}
\eeq
The second line in (\ref{k2}), that 
is generated  by a change of variables in the 
integral ($\t U=-U-2$) while  using (\ref{k1}), also gives a 
relation between $k_1$ and $k_2$,
\beq
k_2=6^{\sqrt{3}/4}k_1~.
\eeq
The solution has the following 
asymptotic behavior. At large distances, $r\gg r_*$ ($U\to -2^-$), we 
derive 
\beq
{P(r)\over r} =-{\t r_{M_2}^2\over r^2}+m_c^2r^2+\dots~,
\eeq
where,
\beq
\t r_{M_2}^2={3\over (3-\sqrt{3})^{2\sqrt{3}}} {m_c^2\over k_2^4}
\approx 0.45\ r_M r_*~,
\eeq
while at short distances, $r\ll r_*$ ($U\to -\infty$), we get
\beq
{P(r)\over r}={r_M\over r}-\alpha_2 m_c^2r^2 
\left({r_*\over r}\right)^{2(\sqrt{3}-1)}+\dots~,
\eeq
where $\alpha_2=-\alpha_1 \approx -0.84$ is, in absolute value, the same 
constant appearing in the conventional 
branch short distance expansion (\ref{short}). Note, however, that the sign of 
the correction to the $4D$ behavior is opposite in the two branches. 

At intermediate  distances, $r_* \ll  r \ll r_c$,  
the potential contains a $5D$ gravitational term that 
is {\em repulsive}, ${\tilde r}^2_M/r^2$. 
This looks like a  5D negative mass. However, this is not an 
asymptotic value of the mass since one can 
only cover the solution in the above  coordinate system till 
$r\sim r_c$ where the dS like horizon is encountered. 
Moreover, in the intermediate regime
$r_* \ll  r \ll r_c$, the  de Sitter term $m_c^2 r^2$ 
in the potential always dominates over the 
${\tilde r}^2_M/r^2$ term suggesting that the effects 
due to the Schwarzschild source are strongly suppressed.

\subsection{Perihelion precession}

{}The deviation from 4D gravity (\ref {potential}) 
gives rise to  the additional perihelion 
precession of circular orbits \cite{Lue1,DGZ} 
(see also \cite {Iorio1} for comprehensive studies
of these and related issues). In a simplest approximation this effect 
is quantified by a fraction of the deviation of the potential 
from its Newtonian form 
\beq
\epsilon \equiv {\Delta \phi \over \phi}\,.
\label{epsilon}
\eeq
This can be  used  to evaluate
an additional perihelion precession of orbits
in the Solar system \cite {Lue1,DGZ}\footnote{Note that in the 
leading order of the relativistic expansion the answer 
is given by the correction to the Newtonian potential, while 
the correction  to the $rr$ component of the metric is 
not important.}. As we discussed in Section 1, the $\epsilon$ 
ratio is somewhat different for the non-perturbative solution
(NPS solution) as compared to the approximate solution (the PS solution ) 
used in Refs. \cite {Lue1,DGZ}. We can easily calculate this difference:
\beq
{\epsilon_{NPS}\over \epsilon_{PS}} \simeq  {|\alpha|\over \sqrt{2}}
\left ({r\over r_*}\right )^\beta \simeq 0.59 
\left ({r\over r_*}\right )^{0.04}\,.
\label{eratio} 
\eeq
The perihelion precession per orbit is
\beq
\Delta \varphi = 2\pi  +  {3\pi r_M\over r} \mp {3\pi |\alpha| \over 4}
\left ( {r\over r_*}\right )^{3/2} \,\left ({r\over r_*}\right )^{0.04}\,.
\label{anprec}
\eeq
The second term on the RHS is the Einstein precession,
and the last term arises due to modification of gravity.
For the PS solution this was first calculated in Refs. 
\cite {Lue1,DGZ}; the solution (\ref {anprec})
is written for the NPS solution and is somewhat different. 

For the Earth-Moon system $r=3.84 \times 10^{10}~cm$ and 
$r^{Earth}_* \simeq 6.59 \times 10^{12}~cm$; 
as a result the ratio in (\ref {eratio})
is approximately $0.48$. Therefore, the predictions of the 
NPS solution for the additional perihelion precession of the
Moon is a factor of two smaller than the predictions of the 
approximate solution. The result of (\ref {anprec}) for the 
additional precession (the last term on the RHS) is 
$\mp 0.7 \times 10^{-12}$ (the plus sign 
for the selfaccelerated branch). This is below  the current 
accuracy of $2.4 \times 10^{-11}$ \cite {lunardata}, but 
could potentially be probed in the near future \cite {adel}
\footnote{An interesting possibility that similar effects 
could leed to seemingly observable increase of the Astronomical Unit
was recently discussed in \cite {Iorio2}.}. 

A similar calculations can be performed for the anomalous Martian 
precession \cite {Lue1,DGZ}. For the Sun-Mars system we get:
\beq
{\epsilon_{NPS}\over \epsilon_{PS}} \simeq 0.59 
 \left ({r_{Sun-Mars}\over r^{Sun}_*}\right )^{0.04}\simeq 0.30\,,
\label{eratiomars} 
\eeq
where we used $r_{Sun-Mars}= 2.28 \times 10^{13}~ cm $   and 
$r^{Sun}_*=4.9 \times  10^{20} ~cm$.  Therefore,  we see that the 
suppression in the NPS result for the precession of the Martian orbit 
is stronger. The additional precession of the Mars orbit 
is $ \sim \mp 1.3\times 10^{-11}$, which should be contrasted 
with a potential accuracy of the Pathfinder mission  
$ \sim 9\times 10^{-11}$.

Last but not least, Lue and Starkman (LS) \cite {Lue1},
found that  the PS solution gives rise
to a correction to the precession {\it rate} 
(additional precession per unit time), 
\beq
\Gamma_{LS} = \mp {3\over 8r_c}\,,
\label{LS}
\eeq
that is universal, i.e., is  independent of  the source.  
The NPS solution, predicts a  weak anomalous  violation of 
the universal Lue-Starkman scaling due to the RHS of 
(\ref {eratio}). The results is  
\beq
\Gamma = \Gamma_{LS} \times 
{|\alpha|\over \sqrt{2}} \left ({r\over r_*}\right )^{0.04}\,.
\label{newrate}
\eeq
This rate depends mildly on the source mass and a separation from it.
The rate is a slowly increasing function or $r$, 
as opposed to the rate due to the second term on the RHS of 
(\ref {anprec}), which is decreasing with growing 
$r$ as $\Gamma_{Einstein}= \sqrt{9r_M^3/8 r^5}$.

\section{Outlook}

In this note we compared the PS \cite {DGP,DDGV,Gruzinov,Lue1,Porrati} 
and NPS \cite {GI} solutions in the DGP model. We emphasized different, 
but interesting  
predictions  that these two solutions make in the observationally 
accessible domain of $r\ll r_*$. These predictions are testable.

As we have also mentioned, there will be important differences in the 
predictions at $r\gg r_*$. These need further detailed studies, 
especially in the context of the structure formation. 
We would expect that both the linear as well as non-linear regimes
of the structure formation will be affected. If the NPS solution 
is the right one, then even at very large scales nonperturbative 
techniques should be used. Moreover, the nonlinear 
regime of the structure formation could be sensitive to, and be able to 
discriminate between, the PS and NPS solutions.

The  same issue  of nonlinear interactions arises in the 
context of strong coupling behavior in the 5D DGP model 
\cite {DDGV,Ratt,Rubakov,Dvali,Nicolis}.  This is 
related to the problem of the UV completion of the quantum 
theory  \cite {Ratt,Rubakov} for which seemingly two different
proposals were put forward in Refs. \cite {Dvali} and \cite {Nicolis}.
It would be interesting to pursue these studies further.
The string theory realizations of brane induced gravity 
of Refs. \cite {Kiritsis,Ignatios,Eman} can be taken as 
a guideline. It would also be interesting 
to understand the NPS solution in terms of the approach 
of Refs. \cite {Ratt,Nicolis}. 

We have not touched upon the issue 
whether the small fluctuations on the selfaccelerated 
branch contain a negative norm state \cite {Ratt,Nicolis},
or not ( see also \cite {Koyama}), and when these 
fluctuations are relevant. Additional investigations 
on this issue are being conducted.

It would also be interesting to look at the Schwarzschild 
solutions in models of large distance modified gravity where 
nonlinear interactions do not exhibit the strong coupling behavior. 
This is the case \cite {GS} in a certain models of brane induced 
gravity in more  than five dimensions \cite {DG,GS}, as well as 
in the  ``dielectric regularization'' of the 5D DGP model 
\cite {PorratiRomb}. Finally we would also point out the constrained 
approach to the 5D  DGP model \cite {GG,SiopsisC,MGU} in which 
case strong interactions also seem to be absent. All the 
above deserves further detailed investigations.

%

\section*{Acknowledgments}

{} We would like to thank Cedric Deffayet, Gia Dvali, Andrei Gruzinov, 
Nemanja Kaloper, Arthur Lue, Roman Scoccimarro, and Glenn Starkman for 
discussions, we also thank Dr. L. Iorio for useful correspondence. 
The work was supported in part by NASA Grant NNGG05GH34G, 
and in part by NSF Grant PHY-0403005.

\end{document}